\documentclass[aps, pra, twocolumn, showpacs, superscriptaddress]{revtex4-1}
\usepackage{amsmath, amssymb, graphicx, dcolumn, subfigure, CJK}
\usepackage{mathptmx, color, hyperref}
\usepackage[all]{hypcap}

\newcommand{\mytitle}{Impact of hollow-atom formation on coherent x-ray scattering at high intensity}

\definecolor{MyDarkGreen}{rgb}{0,0.6,0}
\definecolor{MyDarkBlue}{rgb}{0,0,0.8}
\definecolor{MyDarkRed}{rgb}{0.6,0,0.3}
\hypersetup{breaklinks=true,colorlinks=true,plainpages=true,linktocpage=true,linkcolor=MyDarkBlue,citecolor=MyDarkGreen,urlcolor=MyDarkRed,pdfborder={0 0 0},%
pdfauthor={Sang-Kil Son},%
pdftitle={\mytitle}%
}

\newcommand{\reducedmatrix}[3]{\ensuremath{\left\langle#1\left\lVert#2\right\rVert#3\right\rangle}}

\newcommand{\config}[3]{1s$^{#1}$2s$^{#2}$2p$^{#3}$}

\newlength{\figurewidth}
\begin{document}
\begin{CJK*}{UTF8}{}

\title{\mytitle}

\author{Sang-Kil Son \CJKfamily{mj}(손상길)}
\email{sangkil.son@cfel.de}
\affiliation{Center for Free-Electron Laser Science, DESY, 22607 Hamburg, Germany}

\author{Linda Young}
\email{young@anl.gov}
\affiliation{Argonne National Laboratory, Argonne, Illinois 60439, USA}

\author{Robin Santra}
\email{robin.santra@cfel.de}
\affiliation{Center for Free-Electron Laser Science, DESY, 22607 Hamburg, Germany}
\affiliation{Department of Physics, University of Hamburg, 20355 Hamburg, Germany}

\date{\today}

\begin{abstract}
X-ray free-electron lasers (FELs) are promising tools for structural determination of macromolecules via coherent x-ray scattering.
During ultrashort and ultraintense x-ray pulses with an atomic scale wavelength, samples are subject to radiation damage and possibly become highly ionized, which may influence the quality of x-ray scattering patterns.
We develop a toolkit to treat detailed ionization, relaxation, and scattering dynamics for an atom within a consistent theoretical framework.
The coherent x-ray scattering problem including radiation damage is investigated as a function of x-ray FEL parameters such as pulse length, fluence, and photon energy.
We find that the x-ray scattering intensity saturates at a fluence of $\sim 10^{7}$ photons/\AA$^2$ per pulse, but can be maximized by using a pulse duration much shorter than the time scales involved in the relaxation of the inner-shell vacancy states created.
Under these conditions, both inner-shell electrons in a carbon atom are removed, and the resulting hollow atom gives rise to a scattering pattern with little loss of quality for a spatial resolution $> 1$~\AA.
Our numerical results predict that in order to scatter from a carbon atom 0.1 photons per x-ray pulse, within a spatial resolution of 1.7~\AA, a fluence of $1\times 10^{7}$ photons/\AA$^2$ per pulse is required at a pulse length of 1~fs and a photon energy of 12~keV.
By using a pulse length of a few hundred attoseconds, one can suppress even secondary ionization processes in extended systems.
The present results suggest that high-brightness attosecond x-ray FELs would be ideal for single-shot imaging of individual macromolecules.
\end{abstract}

%

\maketitle
\end{CJK*}

\section{Introduction}\label{sec:intro}

X-ray free-electron lasers (FELs)~\cite{Feldhaus05,Pellegrini04} provide unparalleled peak brightness and open a new era in science and technology, offering many possibilities that have not been conceivable with conventional light sources~\cite{Mancuso10,Piancastelli10,Neutze04}.
The world's first x-ray FEL---the Linac Coherent Light Source (LCLS) at SLAC National Accelerator Laboratory~\cite{Emma10}---has been in operation since 2009, with a photon energy of up to 8.3~keV, up to $2\times10^{12}$~photons per pulse, and a full-width-half-maximum (FWHM) pulse length as short as a few femtoseconds.
The SPring-8 Compact SASE Source (SCSS) at SPring-8~\cite{SCSS} and the European X-ray FEL at DESY~\cite{XFEL} are under construction and are planned to deliver up to 12~keV photon energy with an average brightness 5--500 times higher than LCLS.

One of the prospective applications of x-ray FELs is single-shot imaging of individual macromolecules~\cite{Hajdu00,Gaffney07,Miao08}, which employs coherent x-ray scattering to determine the atomically resolved structure of non-crystallized biomolecules or other nanoparticles~\cite{Miao99,Robinson01,Chapman06,Chapman07,Schmidt08,Barty08,Marchesini08,Bogan08,Nishino09}.
Single-shot imaging becomes possible because the high fluence of a tightly focused x-ray FEL pulse could produce a significant amount of scattered photons from single-molecule samples.
One of the key challenges in single-shot imaging using ultraintense x rays is radiation damage~\cite{Howells09}.
Each target molecule undergoes electronic damage via processes such as photoionization and Auger decay.  The positively charged
atomic ions formed in this way repel each other, thus leading to Coulomb explosion of the target molecule~\cite{Wabnitz02,Saalmann02,Ziaja09a}.
Since the fluence required for single-shot imaging exceeds the conventional damage limit (200 photons/\AA$^2$)~\cite{Henderson95}, these damage effects could degrade the scattering patterns and hinder the determination of the atomic positions in the target molecule.
To suppress the impact of the molecular Coulomb explosion on atomically resolved imaging, one must effectively freeze the atomic motion during the x-ray pulse, requiring a pulse duration of no more than ten femtoseconds~\cite{Solem82,Neutze00}.
Following this idea, several theoretical approaches were used to simulate the radiation damage processes including the movement of the atoms:  
References~\cite{Neutze00,Jurek04,Jurek04a,Bergh04,Faigel05,Jurek08,Jurek09} employed molecular dynamics, Refs.~\cite{Hau-Riege04,Hau-Riege05,Hau-Riege07b,Hau-Riege07c,Hau-Riege08} based their description on a hydrodynamic model, and Refs.~\cite{Ziaja06a,Ziaja08,Ziaja08a,Ziaja09,Ziaja09a} used a kinetic Boltzmann model.
%

For {\em electronic} damage processes, which in the x-ray regime are mainly atom-specific, one may concentrate, to a first
approximation, on the interaction of the x rays with individual atoms.  The dynamics of bound electrons in an isolated atom during an ultraintense x-ray pulse were investigated theoretically in connection with hollow-atom formation~\cite{Moribayashi98,Rohringer07}, x-ray fluorescence~\cite{Moribayashi04,Moribayashi07,Moribayashi08}, and radiation damage~\cite{Kai10}.
It is important to note that even if the atomic motion during an x-ray pulse is negligible, electronic damage dynamics during the x-ray pulse may directly influence x-ray scattering patterns by altering the electronic density in the target~\cite{Hau-Riege07}.  It is, therefore, crucial to understand detailed ionization and relaxation dynamics in individual atoms under ultrashort and ultraintense x-ray pulses.

A series of recent experiments conducted at LCLS revealed how electrons interact with ultraintense, ultrafast x-ray pulses~\cite{Young10,Hoener10,Cryan10,Fang10}.
In the x-ray regime, photoabsorption predominantly affects inner-shell (core) electrons.  If inner-shell photoabsorption is saturated, all inner-shell electrons in a given atom may be removed before Auger decay or other relaxation processes occur~\cite{Moribayashi98,Rohringer07}.  The transient state thus produced is referred to as a hollow atom.
If the pulse length is short enough, the hollow atom retains its core vacancies during the pulse.  Because the x-ray photoabsorption probability is smaller for valence electrons than for inner-shell electrons, hollow-atom formation suppresses further electronic damage.  This effect is called x-ray transparency~\cite{Young10} or frustrated absorption~\cite{Hoener10},
and might be beneficial for single-shot imaging of individual molecules~\cite{Young10,Hoener10}.  
The present paper investigates this idea in detail and provides criteria for the x-ray FEL parameters required.

To treat x-ray--atom interactions, we employ a consistent \textit{ab initio} framework~\cite{Santra09} based on nonrelativistic quantum electrodynamics and perturbation theory.
This x-ray atomic theory has been applied to study x-ray absorption by laser-dressed atoms~\cite{Buth07,Buth07a,Buth08,Glover10} and x-ray scattering from laser-aligned molecules~\cite{Ho08,Ho09a,Pabst10}.
In this paper, we present a practical implementation of this \textit{ab initio} framework as a toolkit to calculate cross sections and rates of x-ray-induced processes for various charge states and electronic configurations of an isolated atom within an approximation to the Hartree--Fock model.
With those parameters, we simulate hollow-atom formation dynamics under ultrashort and ultraintense x-ray pulses by means of time-dependent rate equations~\cite{Rohringer07,Makris09,Young10}.
Then we investigate the effects of hollow-atom formation on coherent x-ray scattering from atomic carbon.

The paper is organized as follows.
In Sec.~\ref{sec:method}, we present theoretical methods to compute cross sections and rates of all electronic damage processes, which are integrated into a set of rate equations.
In Sec.~\ref{sec:results}, we explore coherent x-ray scattering signals influenced by hollow-atom formation and their dependence on x-ray FEL parameters such as pulse length, fluence, and photon energy, and on the spatial resolution of the image.
We also discuss the role of electron impact ionization in molecules.
We conclude with a summary and future perspectives in Sec.~\ref{sec:conclusion}.

\section{Theory and Numerical details}\label{sec:method}
The present toolkit of x-ray atomic processes covers ionization, relaxation (Auger decay and fluorescence), and coherent x-ray scattering.
We assume that inelastically (Compton) scattered photons are energetically distinguishable from elastically (coherently) scattered photons and focus on coherent scattering processes for imaging problems.
Compton scattering contributes to electronic damage, but is negligible in comparison with photoionization for the photon energies under consideration~\cite{Thompson01}.
Shake-up and shake-off processes~\cite{Carlson65,Omar92,Persson01} also make a small contribution to electronic damage and are not included in our model.
We also neglect impact ionization~\cite{Hau-Riege04,Jurek04,Ziaja05,Ziaja06,Kai10}, i.e., secondary ionization in molecules 
induced by photoelectrons and/or Auger electrons via (e,2e) processes.
We will discuss a straightforward strategy to reduce impact ionization in Sec.~\ref{sec:impact_ionization}.
Atomic units are used in this section.

\subsection{Hartree--Fock--Slater model}\label{sec:HFS}
In order to implement the \textit{ab initio} framework~\cite{Santra09}, we use the Hartree--Fock--Slater (HFS) model~\cite{Slater51,Herman63}, which employs a local density approximation to the exact exchange interaction.
The effective one-electron (mean-field) Schr\"odinger equation to be solved is
\begin{equation}\label{eq:SE}
\left[ -\frac{1}{2} \nabla^2 + V(\mathbf{r}) \right] \psi(\mathbf{r}) = \varepsilon \psi(\mathbf{r}).
\end{equation}
Here the potential is given by
\begin{equation}\label{eq:potential}
V(\mathbf{r}) = - \frac{Z}{r} + \int \! \frac{\rho(\mathbf{r}')}{| \mathbf{r} - \mathbf{r}' |} \; d^3r' + V_\text{x}(\mathbf{r}),
\end{equation}
where $Z$ is the nuclear charge and the electronic density $\rho(\mathbf{r})$ is given by
\begin{equation}
\rho(\mathbf{r}) = \sum_i^{N_\text{elec}} \psi_i^{\dag}(\mathbf{r})\psi_i(\mathbf{r}),
\end{equation}
where $i$ is the spin--orbital index and $N_\text{elec}$ is the number of electrons.
The exchange term is approximated by the Slater exchange potential~\cite{Slater51},
\begin{equation}\label{eq:V_x}
V_\text{x}(\mathbf{r}) = - \frac{3}{2} \left[ \frac{3}{\pi} \rho(\mathbf{r}) \right]^{1/3}.
\end{equation}
In addition, the potential includes the Latter tail correction~\cite{Latter55} to obtain the proper long-range potential for both occupied and unoccupied orbitals, i.e., we put $V(\mathbf{r}) = - (Z' + 1)/r$ if the right-hand side of Eq.~(\ref{eq:potential}) is less negative than $-(Z' + 1)/r$, where $Z' = Z - N_\text{elec}$ is the effective charge of the system.

After angular momentum averaging, the problem becomes spherically symmetric, and each solution of Eq.~(\ref{eq:SE}) can be expressed in terms of the product of a radial wave function and a spherical harmonic.
For example, a bound-state spatial orbital with quantum numbers $(n,l,m)$ may be written as
\begin{equation}
\psi_{nlm}(\mathbf{r}) = \frac{P_{nl}(r)}{r} Y^m_l(\theta,\phi).
\end{equation}
For the bound states, the radial wave function $P_{nl}(r)$ is accurately solved by the generalized pseudospectral method~\cite{Yao93a,Tong97a} on a nonuniform grid.
For the continuum states, $P_{\varepsilon l}(r)$ is numerically solved by the fourth-order Runge--Kutta method for a given energy $\varepsilon$ on a uniform grid~\cite{Cooper62,Manson68}.
To evaluate integrals involving both bound and continuum states, we use spline interpolation to map the bound-state orbitals from the nonuniform grid to the denser uniform grid employed for the continuum states.

For the bound-state calculations, the theoretical procedure presented here is identical to the Herman--Skillman code~\cite{Herman63}, which has been widely used in atomic physics, but the numerical part of the present toolkit utilizes a different grid method with the following advantages.
Firstly, it is easy to control convergence with respect to the grid parameters.
Secondly, we can avoid truncation of the maximum radius internally imposed by the Herman--Skillman code, which causes numerical instability for photoabsorption cross section calculations.
Lastly, the matrix eigenvalue problem is solved by a modern linear algebra package~\cite{LAPACK}.
For the present calculations, we use 200 grid points and a maximum radius of 50~a.u.\ for the bound states (nonuniform grids) and a radial step size of 0.001~a.u.\ for the continuum states (uniform grids) to achieve machine accuracy for cross sections and rates.

We calculate all cross sections and rates for all possible electronic configurations.
For example, a neutral carbon atom has a \config{2}{2}{2} ground configuration, so the number of all configurations that can be formed by removing 0, 1, 2, 3, 4, 5, or all 6 electrons from the occupied orbitals is $1+3+6+7+6+3+1 = 27$.
Note that we perform a separate HFS calculation for each configuration.
In other words, the orbitals are optimized in the presence of core and/or valence vacancies.  Thus, orbital relaxation for the core-hole configurations is automatically included, a strategy that is known to be in good agreement with multiconfigurational self-consistent-field calculations~\cite{Hau-Riege07}.

\subsection{X-ray absorption process}\label{sec:photionization}
The cross section for ionizing an electron in the $i$th subshell by absorbing an x-ray photon with energy $\omega$ is given by~\cite{Santra09}
\begin{widetext}
\begin{equation}
\sigma_\text{P}(i,\omega) = \frac{4}{3} \alpha \pi^2 \omega
N_i \sum_{l_j = |l_i-1|}^{l_i+1} \frac{l_>}{2 l_i + 1} 
\left| \int_0^\infty P_{n_i l_i}(r) P_{\varepsilon l_j}(r) \; r \; dr \right|^2,
\end{equation}
\end{widetext}
where $\alpha$ is the fine-structure constant, $N_i$ is the occupation number of the $i$th subshell, $l_> = \max( l_i, l_j )$, and $\varepsilon = \omega - E_i$ is the photoelectron energy.
Here, $E_i$ is the ionization energy of the $i$th subshell ($E_i = -\varepsilon_i$) by Koopmans' theorem~\cite{Koopmans34}, which is approximately valid in the HFS model.
The orbital energy $\varepsilon_i$ and the radial wave functions $P_{n_i l_i}(r)$ and $P_{\varepsilon l_j}(r)$ are calculated for a given electronic configuration.
We do not consider orbital hole alignment after ionization by linearly-polarized x-ray pulses and, hence, 
assume that the density of bound electrons remains spherically symmetric throughout.

Table~\ref{table:pcs} shows x-ray absorption cross sections for all configurations of carbon (except the bare nucleus) at photon energies of 8~keV and 12~keV, respectively.  The photon energy range considered here is consistent with that available at current and future x-ray FELs~\cite{Emma10,SCSS,XFEL}.
Note that the present total cross sections are in agreement with those in the literature~\cite{Verner93} to within less than 10\%.

The impact of orbital relaxation is evident in the subshell cross sections shown in Table~\ref{table:pcs}.
For instance, the 2s subshell cross section for the \config{1}{2}{2} and \config{0}{2}{2} configurations differ by 40\% and 90\%, respectively, from the 2s subshell cross section for the ground configuration of neutral carbon.

As may be seen in Table~\ref{table:pcs}, the cross sections for the 1s and 2s subshells are much greater than the 2p subshell cross section, because $\sigma_\text{P}$ is proportional to $\omega^{-l-7/2}$ in the high energy limit~\cite{Bethe08}.
Therefore, absorption of linearly polarized x rays does not, in general, induce any orbital hole alignment.
This justifies our assumption of spherically symmetric electron densities.

\begin{table*}
\caption{\label{table:pcs}%
X-ray absorption cross sections ($\sigma_\text{P}$) for various configurations of carbon at 8~keV and 12~keV.
}
\begin{ruledtabular}
\begin{tabular}{ccdddddd}
 &         
 & \multicolumn{3}{c}{$\sigma_\text{P}$ (10$^{-8}$ a.u.) at 8~keV}
 & \multicolumn{3}{c}{$\sigma_\text{P}$ (10$^{-8}$ a.u.) at 12~keV}
\\
\cline{3-5} \cline{6-8}
Charge & Configuration 
 & \multicolumn{1}{c}{1s} 
 & \multicolumn{1}{c}{2s} 
 & \multicolumn{1}{c}{2p}
 & \multicolumn{1}{c}{1s} 
 & \multicolumn{1}{c}{2s} 
 & \multicolumn{1}{c}{2p}
\\
\hline
$+0$   & \config{2}{2}{2} & 287 & 14.8 & 0.0897 & 77.8 & 4.05 & 0.0164 \\
\hline                                                               
$+1$   & \config{1}{2}{2} & 155 & 20.8 & 0.219  & 41.9 & 5.70 & 0.0399 \\
      & \config{2}{1}{2} & 287 & 8.28 & 0.118  & 78.0 & 2.27 & 0.0214 \\
      & \config{2}{2}{1} & 287 & 16.6 & 0.0590 & 77.9 & 4.53 & 0.0107 \\
\hline                                                               
$+2$   & \config{0}{2}{2} & -   & 27.9 & 0.387  & -    & 7.61 & 0.0699 \\
      & \config{1}{1}{2} & 155 & 11.5 & 0.258  & 42.0 & 3.16 & 0.0474 \\
      & \config{1}{2}{1} & 155 & 23.4 & 0.132  & 42.0 & 6.41 & 0.0243 \\
      & \config{2}{0}{2} & 288 & -    & 0.145  & 78.2 & -    & 0.0266 \\
      & \config{2}{1}{1} & 288 & 9.25 & 0.0737 & 78.2 & 2.53 & 0.0134 \\
      & \config{2}{2}{0} & 288 & 18.6 & -      & 78.1 & 5.10 & -      \\
\hline                                                               
$+3$   & \config{0}{1}{2} & -   & 15.1 & 0.435  & -    & 4.13 & 0.0790 \\
      & \config{0}{2}{1} & -   & 31.1 & 0.221  & -    & 8.48 & 0.0397 \\
      & \config{1}{0}{2} & 156 & -    & 0.300  & 42.1 & -    & 0.0548 \\
      & \config{1}{1}{1} & 156 & 12.8 & 0.153  & 42.2 & 3.51 & 0.0280 \\
      & \config{1}{2}{0} & 156 & 26.0 & -      & 42.3 & 7.11 & -      \\
      & \config{2}{0}{1} & 289 & -    & 0.0885 & 78.4 & -    & 0.0165 \\
      & \config{2}{1}{0} & 289 & 10.8 & -      & 78.5 & 2.93 & -      \\
\hline                                                               
$+4$   & \config{0}{0}{2} & -   & -    & 0.454  & -    & -    & 0.0841 \\
      & \config{0}{1}{1} & -   & 16.6 & 0.246  & -    & 4.54 & 0.0442 \\
      & \config{0}{2}{0} & -   & 33.8 & -      & -    & 9.23 & -      \\
      & \config{1}{0}{1} & 156 & -    & 0.180  & 42.2 & -    & 0.0323 \\
      & \config{1}{1}{0} & 157 & 14.7 & -      & 42.4 & 4.03 & -      \\
      & \config{2}{0}{0} & 284 & -    & -      & 76.3 & -    & -      \\
\hline                                               
$+5$   & \config{0}{0}{1} & -   & -    & 0.285  & -    & -    & 0.0522 \\
      & \config{0}{1}{0} & -   & 18.0 & -      & -    & 4.91 & -      \\
      & \config{1}{0}{0} & 156 & -    & -      & 42.2 & -    & -      \\
\end{tabular}
\end{ruledtabular}
\end{table*}

\subsection{Auger decay process}\label{sec:Auger}
The Auger decay rate that an electron from the $j$th subshell fills the $i$th subshell and another electron from the $j'$th subshell is ejected into the continuum may be written as~\cite{Santra09,Bhalla73}
\begin{widetext}
\begin{equation}\label{eq:Auger}
\Gamma_\text{A}(i,j j') = \pi \frac{N^\text{H}_i N_{j j'}}{2l_i + 1} \sum_{L=|l_j - l_{j'}|}^{l_j+l_{j'}} \sum_{S=0}^{1}  \; \; \sum_{l_{i'}} (2L+1) (2S+1) | M_{LS}(j,j',i,i') |^2,
\end{equation}
where $i'$ indicates the continuum state with Auger electron energy $\varepsilon = E_i - E_j - E_{j'}$, $N^\text{H}_i$ is the number of holes in the $i$th subshell, and 
\begin{equation}
N_{j j'} = 
\begin{cases}
\frac{ N_j N_{j'} }{ ( 4 l_j + 2 ) ( 4 l_{j'} + 2 ) } & 
 \text{for inequivalent electrons},
\\
\frac{ N_j ( N_j - 1 ) }{ ( 4 l_j + 2 ) ( 4 l_j + 2 - 1 ) } & 
 \text{for equivalent electrons}.
\end{cases}
\end{equation}
Here, averaging schemes over initial and final states to compute transition rates are adopted from Refs.~\cite{Walters71,Walters71b,Bhalla73}.

The matrix element $M_{LS}$ is given by
\begin{equation}
M_{LS}(j,j',i,i') 
= \tau (-1)^{L + l_j + l_{i'}} \sum_K \left[ R_K(j,j',i,i') A_K(j,j',i,i') + (-1)^{L+S} R_K(j',j,i,i') A_K(j',j,i,i') \right] 
\end{equation}
\end{widetext}
where $\tau = 1 / \sqrt{2}$ if $j$ and $j'$ are equivalent electrons and $\tau=1$ otherwise.
$A_K$ is a coefficient related to 3j- and 6j-symbols~\cite{Zare88},
\begin{equation}
A_K(j,j',i,i') = \reducedmatrix{l_{i}}{C_K}{l_{j}} \reducedmatrix{l_{i'}}{C_K}{l_{j'}}
\begin{Bmatrix}
l_{i}  & l_{i'} & L \\
l_{j'} & l_{j}  & K
\end{Bmatrix},
\end{equation}
where
\begin{equation}
\reducedmatrix{l}{C_K}{l'} = (-1)^l \sqrt{ (2l+1) (2l'+1) }
\begin{pmatrix}
l' & K & l \\
0  & 0 & 0
\end{pmatrix},
\end{equation}
and $R_K$ is a double radial integral defined as
\begin{widetext}
\begin{equation}
R_K(j,j',i,i') = \int_0^\infty \! \! \int_0^\infty 
P_{n_j l_j}(r_1) P_{n_{j'} l_{j'}}(r_2) 
\frac{r_<^K}{r_>^{K+1}} 
P_{n_i l_i}(r_1) P_{\varepsilon l_{i'}}(r_2) \; d r_1 \; d r_2.
\end{equation}
\end{widetext}

Table~\ref{table:rates} lists the Auger rates computed using Eq.~(\ref{eq:Auger}) for various configurations of carbon.
$K L_1 L_1$, $K L_1 L_{23}$, and $K L_{23} L_{23}$ represent $\Gamma_\text{A}$(1s,2s2s), $\Gamma_\text{A}$(1s,2s2p), and $\Gamma_\text{A}$(1s,2p2p), respectively.
For the 1s hole configuration, the present results are in good agreement with other theoretical results~\cite{McGuire69,Walters71,Moribayashi08}.
Note that Ref.~\cite{Moribayashi08} includes Auger and fluorescence rates for all possible configurations of carbon computed by Cowan's atomic structure code~\cite{Cowan81}.  
The decay rates in Ref.~\cite{Moribayashi08} are in fair agreement, to within about 40\%, with the present results.
We note that the experimental Auger lifetime for a free carbon ion, averaged over the subset of all doublet states in the \config{1}{2}{2} configuration, is 7.3~fs~\cite{Schlachter04}, and possibly fortuitously agrees somewhat better with our calculated value ($\approx$10~fs) than with Ref.~\cite{Moribayashi08} ($\approx$14~fs).

\subsection{Fluorescence process}\label{sec:fluorescence}
The fluorescence rate for the electric dipole transition of an electron from the $j$th subshell to a hole in the $i$th subshell is given by~\cite{Santra09,Bhalla73}
\begin{widetext}
\begin{equation}
\Gamma_\text{F}(i,j) = \frac{4}{3} \alpha^3 ( I_i - I_j )^3 
\frac{N^\text{H}_i N_j}{4 l_j + 2} \cdot
\frac{l_>}{2 l_i + 1} \left| \int_0^\infty P_{n_i l_i}(r) P_{n_j l_j}(r) \; r \; dr \right|^2.
\end{equation}
\end{widetext}
The last two columns in Table~\ref{table:rates} show the x-ray fluorescence rates and yields [$=\Gamma_\text{F}/(\Gamma_\text{F}+ \sum \Gamma_\text{A})$] for various configurations of carbon,
where $K\alpha$ indicates $\Gamma_\text{F}$(1s,2p).
For the 1s hole configuration, the x-ray fluorescence yield is compared with experimental data~\cite{Hubbell94}.
In light atoms like carbon, the x-ray fluorescence yield is generally small.  Note, however, that x-ray fluorescence is 
the only decay process available in C$^{4+}$ \config{1}{0}{1} and C$^{5+}$ \config{0}{0}{1}.
All fluorescence rates are included in our damage dynamics model for completeness.

\begin{table*}
\caption{\label{table:rates}%
Auger rates ($\Gamma_\text{A}$) and fluorescence rates ($\Gamma_\text{F}$) for various configurations of carbon.
For the 1s hole configuration, the present values are compared with other approaches.  A: Herman--Skillman code~\cite{McGuire69}, B: Cowan code~\cite{Moribayashi08}, C: semi-empirical method~\cite{Walters71}, and EXP: experimental data~\cite{Hubbell94}.
}
\begin{ruledtabular}
\begin{tabular}{ccddddc}
 &         
 & \multicolumn{3}{c}{$\Gamma_\text{A}$ (10$^{-3}$ a.u.)} 
 & \multicolumn{1}{c}{$\Gamma_\text{F}$ (10$^{-5}$ a.u.)} 
 & \multicolumn{1}{c}{Fluorescence}
 \\
\cline{3-5} \cline{6-6}
Charge & Configuration 
 & \multicolumn{1}{c}{$K L_1 L_1$} 
 & \multicolumn{1}{c}{$K L_1 L_{23}$} 
 & \multicolumn{1}{c}{$K L_{23} L_{23}$} 
 & \multicolumn{1}{c}{$K \alpha$} 
 & \multicolumn{1}{c}{yield}
 \\
\hline
$+1$  & \config{1}{2}{2} &       &       &       &       & \\
      & Present          & 0.961 & 0.970 & 0.439 & 0.836 & 0.0035\\
      & A                & 0.929 & 0.987 & 0.435 & 0.824 & 0.0035\\
      & B                & 0.680 & 0.697 & 0.392 & 0.651 & 0.0037\\
      & C                & 0.857 & 0.824 & 0.378 & 0.486 & 0.0024\\
      & EXP              & -     & -     & -     & -     & 0.0026\\
\hline                                                     
$+2$  & \config{0}{2}{2} & 2.89  & 3.33  & 1.75  & 2.74  & 0.0034 \\
      & \config{1}{1}{2} & -     & 0.602 & 0.574 & 0.975 & 0.0082 \\
      & \config{1}{2}{1} & 1.18  & 0.620 & -     & 0.498 & 0.0028 \\
\hline                                                     
$+3$  & \config{0}{1}{2} & -     & 1.98  & 2.13  & 3.04  & 0.0073 \\
      & \config{0}{2}{1} & 3.46  & 1.99  & -     & 1.55  & 0.0028 \\
      & \config{1}{0}{2} & -     & -     & 0.703 & 1.11  & 0.0155 \\
      & \config{1}{1}{1} & -     & 0.370 & -     & 0.569 & 0.0151 \\
      & \config{1}{2}{0} & 1.39  & -     & -     & -     & $-$    \\
\hline                                                     
$+4$  & \config{0}{0}{2} & -     & -     & 2.59  & 3.27  & 0.0125 \\
      & \config{0}{1}{1} & -     & 1.15  & -     & 1.69  & 0.0145 \\
      & \config{0}{2}{0} & 3.91  & -     & -     & -     & $-$    \\
      & \config{1}{0}{1} & -     & -     & -     & 0.648 & 1.0000 \\
\hline                                             
$+5$  & \config{0}{0}{1} & -     & -     & -     & 1.97  & 1.0000 \\
\end{tabular}
\end{ruledtabular}
\end{table*}

\subsection{Rate equations for ionization and relaxation dynamics}\label{sec:rate_eq}
To simulate electronic damage dynamics in intense x-ray pulses, we use the rate equation approach that has been successfully used to describe x-ray-induced multiple ionization~\cite{Rohringer07,Makris09,Young10}.
The transitions among all possible electronic configurations $\lbrace I \rbrace$ of a given atom are described by 
a set of coupled rate equations of the form 
\begin{equation}\label{eq:rate_equation}
\frac{d}{dt} P_I(t) = \sum_{I' \neq I}^\text{all config.} \left[ \Gamma_{I' \rightarrow I} P_{I'}(t) - \Gamma_{I \rightarrow I'} P_I(t) \right],
\end{equation}
where $P_I$ is the population of the $I$th configuration, and $\Gamma_{I \rightarrow I'}$ is the rate for transitions from the configuration $I$ to the configuration $I'$.
Here $\Gamma$ can be either a time-independent Auger or fluorescence rate, or a time-dependent photoionization rate given by $\sigma_\text{P} J(t)$, where $J(t)$ is the photon flux of the x-ray pulse at a given time $t$.
In our calculations on carbon, all configurations connected by the photoionization, Auger decay, and x-ray fluorescence processes listed in Tables~\ref{table:pcs} and \ref{table:rates} are included, and the corresponding rate equations are numerically solved using the fourth-order Runge--Kutta method.  We assume that the temporal shape of the x-ray pulse is Gaussian.  
In the regime considered here, the spiky structure of the individual pulses generated by x-ray FELs such as the LCLS is largely irrelevant~\cite{Rohringer07}.

\subsection{Coherent x-ray scattering process}\label{sec:scattering}
The coherent x-ray scattering form factor for a given electronic density $\rho(\mathbf{r})$ is given by~\cite{Santra09}
\begin{equation}\label{eq:cff_orig}
f^0(\mathbf{Q}) = \int \rho(\mathbf{r}) \, e^{i \mathbf{Q} \cdot \mathbf{r}} \; d^3r,
\end{equation}
where $\mathbf{Q}$ is the photon momentum transfer.  We assume that the atomic electron density is spherically symmetric.
Then the atomic form factor depends only on the magnitude of the momentum transfer, so Eq.~(\ref{eq:cff_orig}) may be simplified to
\begin{equation}\label{eq:cff}
f^0(Q) = 4 \pi \int_0^\infty r^2 \rho(r) \frac{\sin( Q r )}{Q r} \; dr,
\end{equation}
where $Q = | \mathbf{Q} | = 2 \alpha \omega \sin( \theta / 2 )$ and $\theta$ is the polar angle of the momentum of the scattered photon with respect to the propagation axis of the incoming x rays.

For the unpolarized x rays, the differential cross section for coherent scattering is given by
\begin{equation}\label{eq:sigma_S}
\frac{d \sigma_\text{S}}{d \Omega} = \alpha^4 | f^0(Q) |^2 \frac{1 + \cos^2 \theta}{2},
\end{equation}
and for linearly polarized x rays, the differential cross section is given by
\begin{equation}\label{eq:sigma_S_LP}
\frac{d \sigma_\text{S}}{d \Omega} = \alpha^4 | f^0(Q) |^2 ( 1 - \cos^2 \phi \sin^2 \theta ),
\end{equation}
where $\phi$ is the azimuthal angle of the scattered photon momentum with respect to the x-ray propagation and polarization axes.
This differential cross section gives the x-ray scattering pattern one would obtain for a fixed electronic configuration.
From measurement of the x-ray scattering pattern, one can retrieve electronic density information.  
Based on the discrete Fourier transform relationship between real space and $Q$ space, the smallest length in real space (the resolution $d$) corresponds to the largest length in $Q$ space ($Q_\text{max}$), while the largest length in real space (the object size $D$) corresponds to the smallest length in $Q$ space (the pixel size $\Delta Q$).
In other words, the spatial resolution desired determines the photon momentum transfer up to which statistically significant scattering data must be available, $Q_\text{max} = 2\pi / d$;
the size of the object in real space determines the maximum pixel size permitted in $Q$ space, $\Delta Q = 2\pi / D$.
For a purely atomic target (the case considered here), the object size is close to the desirable spatial resolution, so all relevant information in momentum space is captured in one pixel.
The number of photons scattered into that pixel is proportional to the integral of the differential cross section over the solid angle $\Omega$ up to the desired resolution.
We note that the integrals of Eqs.~(\ref{eq:sigma_S}) and (\ref{eq:sigma_S_LP}) over $\Omega$ are identical.

During exposure to an ultraintense x-ray pulse, the atomic electron density is dynamically modified, as a consequence of x-ray-induced processes.  
This makes it necessary to introduce a suitably averaged, time-dependent differential scattering cross section, 
\begin{equation}\label{eq:sigma(t)}
\frac{d \sigma_\text{S}}{d \Omega}(t) = \sum_I^\text{all config.} P_I(t) \frac{d \sigma_\text{S}}{d \Omega} \Big\vert_I,
\end{equation}
where $P_I(t)$ is the population of the $I$th configuration, which is obtained as a solution of the rate equations in 
Eq.~(\ref{eq:rate_equation}).  The differential scattering cross section for the $I$th configuration is evaluated from the form factor of Eq.~(\ref{eq:cff}) using the density $\rho(r)$ calculated from the orbitals optimized for the $I$th configuration.
In this way, we incorporate electronic damage dynamics into simulations of coherent x-ray scattering at high intensity.

\subsection{Measurement of the strength and the quality of scattering signals}\label{sec:scattering_signals}
To describe the strength of the x-ray scattering signals, we compute the number of scattered photons ($N_\text{S}$) by integrating over the Ewald sphere~\cite{Als-Nielsen01}, limited to the desired spatial resolution length $d$,
\begin{equation}
N_\text{S}(d) = \int\limits_{Q < Q_\text{max}(d)} \!\! \left[ \int_{-\infty}^{\infty} \! J(t) \frac{d \sigma_\text{S}}{d \Omega}(t) \; dt \right] d\Omega,
\end{equation}
where $J(t)$ is the incident photon flux at a given time $t$, 
and the time-dependent differential scattering cross section is defined in Eq.~(\ref{eq:sigma(t)}).
Here $Q_\text{max}$ is determined by the spatial resolution length, $d = 2\pi / Q_\text{max}$.

\begin{widetext}
To measure the quality of the x-ray scattering patterns, we employ a modified $R$-factor expression with an explicit dependence on the spatial resolution~\cite{Hau-Riege05},
\begin{equation}\label{eq:R-factor}
R(d) = \int\limits_{Q < Q_\text{max}(d)} \!\! \left|
\frac{ \sqrt{ N_\text{real}(\Omega) } }
    { \int\limits_{Q' < Q_\text{max}(d)} \!\!
      \sqrt{ N_\text{real}(\Omega') } \; d\Omega' } 
- \frac{ \sqrt{ N_\text{ideal}(\Omega) } }
      { \int\limits_{Q' < Q_\text{max}(d)} \!\! 
        \sqrt{ N_\text{ideal}(\Omega') } \; d\Omega' } \right| d\Omega,
\end{equation}
\end{widetext}
where 
\begin{equation}
N_\text{real}(\Omega) = \frac{d N_\text{S}}{d \Omega} = \int_{-\infty}^{\infty} \! J(t) \frac{d \sigma_\text{S}}{d \Omega}(t) \; dt
\end{equation}
is the number of photons (per unit solid angle) scattered from the sample undergoing x-ray-induced electronic damage,
and 
\begin{equation}
N_\text{ideal}(\Omega) = \left( \int_{-\infty}^{\infty} \! J(t) \; dt \right) \frac{d \sigma_\text{S}}{d \Omega} \Big\vert_\text{neutral} = F \, \frac{d \sigma_\text{S}}{d \Omega} \Big\vert_\text{neutral}
\end{equation}
is the number of photons (per unit solid angle) scattered from the undamaged sample, which is given by the fluence ($F$) times the differential cross section of the undamaged sample.
In our case, the undamaged sample is the neutral carbon atom in its ground configuration.

\section{Results and discussion}\label{sec:results}
We investigate high-intensity coherent x-ray scattering including electronic damage dynamics with x-ray parameters achievable using x-ray FELs~\cite{Emma10,SCSS,XFEL}.
The FWHM pulse length in our calculations varies from 1~fs to 120~fs, the pulse envelope being a Gaussian.
The number of incident photons varies from $10^{9}$ to $10^{15}$, and the beam size used is 100~nm$\times$100~nm, corresponding to a fluence ranging from $10^3$ to $10^9$ photons/\AA$^2$.  
The photon energy is chosen as 8~keV and 12~keV, respectively.
With the Gaussian envelope and these fluences, the peak intensity ranges from $1\times10^{17}$ to $1\times10^{23}$~W/cm$^2$, for a photon energy of 8~keV and a pulse length of 120~fs.

\subsection{Atomic form factors for core hole states}\label{sec:form_factor}
\begin{figure}
\centering
\includegraphics[width=\figurewidth]{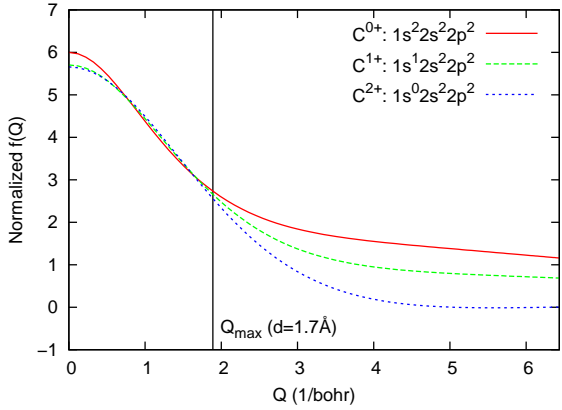}
\caption{(Color online) Normalized atomic form factor for different electronic configurations.}
\label{fig:cff}
\end{figure}

In order to examine variations of x-ray scattering patterns for different electronic configurations, especially for core-hole configurations created via photoabsorption, we calculate atomic form factors for the filled core (neutral: \config{2}{2}{2}), the single-core-hole (\config{1}{2}{2}), and the double-core-hole (\config{0}{2}{2}) configurations (see Fig.~\ref{fig:cff}).
To facilitate a direct comparison between the three different charge states, the form factors in Fig.~\ref{fig:cff} are normalized in the same fashion as used in the $R$-factor expression in Sec.~\ref{sec:scattering_signals}.
Specifically, the normalization factor is 
\begin{equation}
\left.
\int\limits_{Q < Q_\text{max}(d)} \!\! \sqrt{ N_\text{neutral}(\Omega) } \; d\Omega
\middle/
\int\limits_{Q < Q_\text{max}(d)} \!\! \sqrt{ N_I(\Omega) } \; d\Omega,
\right.
\end{equation}
where $N_I(\Omega) = d\sigma_\text{S} / d\Omega$ for a given configuration $I$.
We keep $d=1.7$~\AA\ fixed as a desirable resolution for further analysis.
Within this resolution, i.e., for $Q$ up to $Q_\text{max}$, the shapes of the three normalized form factors 
are quite similar to each other.  
The computed $R$-factors for the single-core-hole and double-core-hole configurations with respect to the ground configuration of the neutral atom are 1.7\% and 2.6\%, respectively.  This fact indicates that core-hole formation causes little degradation of the quality of the x-ray scattering pattern.
We will discuss the optimal resolution minimizing the $R$-factor in Sec.~\ref{sec:resolution}.

\subsection{Hollow-atom formation in ultrashort and ultraintense x-ray pulses}\label{sec:charge}
The time-averaged charge weighted by the normalized pulse envelope provides a simple measure of electronic 
damage during the x-ray pulse~\cite{Kai10,Hau-Riege07}.
The time-averaged population of the $I$th configuration is given by
\begin{equation}\label{eq:time-averaged_population}
\bar{P}_I = \int_{-\infty}^{\infty} \! P_I(t) f(t) \; dt,
\end{equation}
where $P_I(t)$ is the time-dependent population of the $I$th configuration, and $f(t)$ is the normalized x-ray pulse envelope.  
Then the time-averaged charge is given by
\[
\bar{Z} = \sum_I^\text{all config.} Z_I \bar{P}_I,
\]
where $Z_I$ is the charge corresponding to the $I$th configuration.

\begin{figure*}
\centering
\subfigure[\ 8~keV]{\includegraphics[width=\figurewidth]{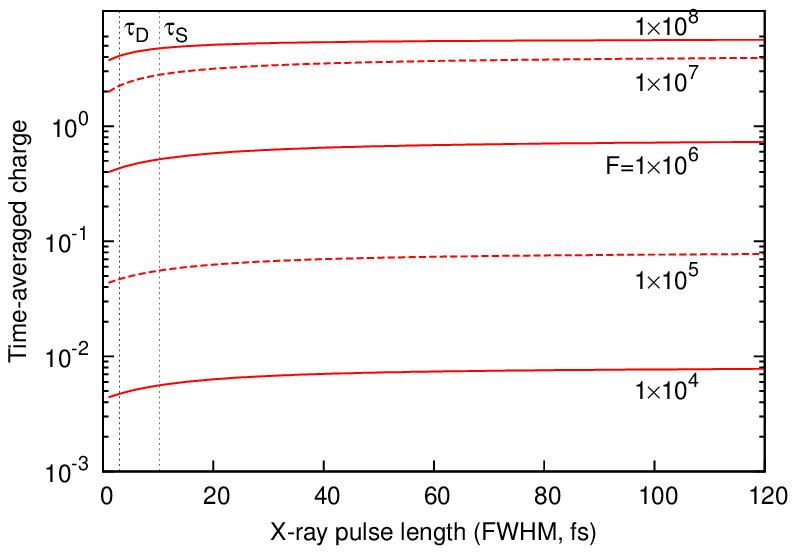}}
\subfigure[\ 12~keV]{\includegraphics[width=\figurewidth]{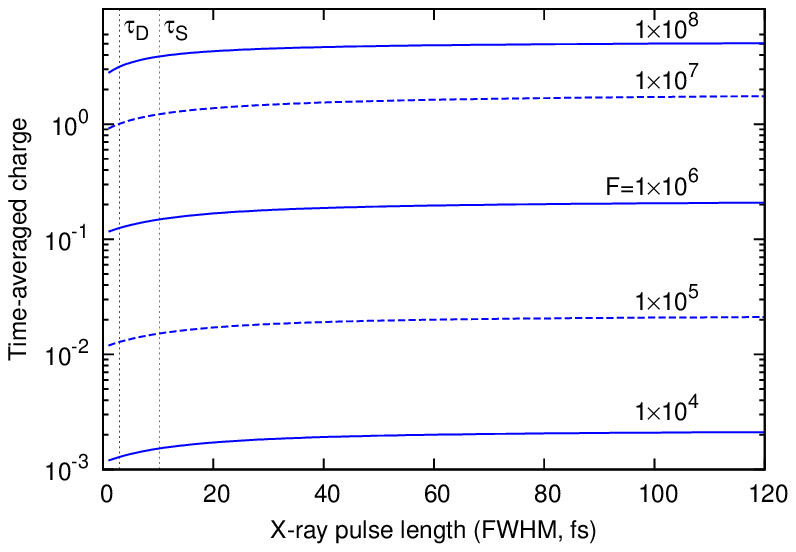}}
\caption{(Color online) Time-averaged charge weighted by the normalized pulse envelope for 8~keV and 12~keV photon energies.  Each line corresponds to a different fluence $F$ (photons/\AA$^2$).
The vertical lines labeled by $\tau_\text{S}$ and $\tau_\text{D}$ indicate the single-core-hole and double-core-hole lifetimes, 
respectively.}
\label{fig:charge}
\end{figure*}

Figure~\ref{fig:charge} shows the time-averaged charge of atomic carbon as a function of the pulse length for several fluences.
When the pulse length is short enough to compete with core-hole lifetimes as marked in Fig.~\ref{fig:charge} ($\tau_\text{S} \approx 10$~fs for the single-core-hole configuration and $\tau_\text{D} \approx 3$~fs for the double-core-hole configuration), $\bar{Z}$ starts to decrease.  This reduction of electronic damage is the signature of x-ray transparency~\cite{Young10} or frustrated absorption~\cite{Hoener10}, which may be understood as follows.
Photoionization of a core electron initiates electronic damage.
If the x-ray pulse length is long enough for Auger decay to occur during the pulse, then a valence electron fills the core vacancy.  Eventually, many electrons can be stripped off in a sequence of core photoionization and Auger decay steps.
Note that in this limit, the effective x-ray absorption cross section remains essentially constant throughout the pulse.
For instance, the x-ray absorption cross sections for the configurations \config{2}{2}{2} (neutral ground configuration) and \config{2}{2}{0} (the dominant Auger decay channel for the single-hole configuration) are almost identical.
As Fig.~\ref{fig:charge} illustrates, the time-averaged charge is practically independent of the pulse duration (and, therefore, independent of the peak intensity) for pulse lengths much longer than 10 fs.
For pulses shorter than 10 fs, there are two stages of x-ray transparency or frustrated absorption.  
In the first stage, if the pulses are still longer than the double-core-hole lifetime ($\tau_\text{D} \approx 3$~fs) and only single-core-hole production is saturated, the effective x-ray absorption cross section drops by a factor of 1.7 relative to the neutral atom (cf.\ Table~\ref{table:pcs}).  
The second stage becomes accessible for pulse durations shorter than the double-core-hole lifetime.  In this case, by saturating hollow-atom formation, the effective x-ray absorption cross section drops further by a factor of 6.2, i.e., relative to the neutral atom the effective x-ray absorption cross section drops by almost a factor of 11 (cf.\ Table~\ref{table:pcs}).
The somewhat counterintuitive consequence of this is that by decreasing the pulse duration and, thereby, increasing the peak intensity, the time-averaged charge $\bar{Z}$ can be minimized, as shown in Fig.~\ref{fig:charge}.

We compare the time-averaged charge for two different photon energies in Fig.~\ref{fig:charge}: (a) 8~keV and (b) 12~keV.
For a given fluence, the time-averaged charge at 8~keV is higher than that at 12~keV, because the photoabsorption cross section at 8~keV is about 4 times higher than that at 12~keV (Table~\ref{table:pcs}).  
This is also expected based on the scaling behavior of the photoabsorption cross section in the high energy limit, $\sigma_\text{P} \propto \omega^{-7/2}\ (l=0)$~\cite{Bethe08}.
Since a higher photon energy induces less electronic damage, it has an advantage with respect to the $R$-factor for x-ray scattering, which will be discussed in Sec.~\ref{sec:resolution}.

\begin{figure*}
\centering
\subfigure[\ 8~keV]{\includegraphics[width=\figurewidth]{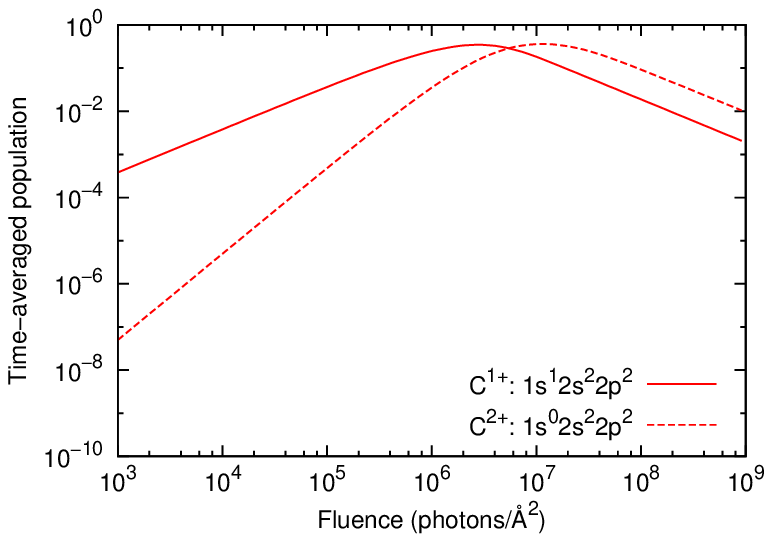}}
\subfigure[\ 12~keV]{\includegraphics[width=\figurewidth]{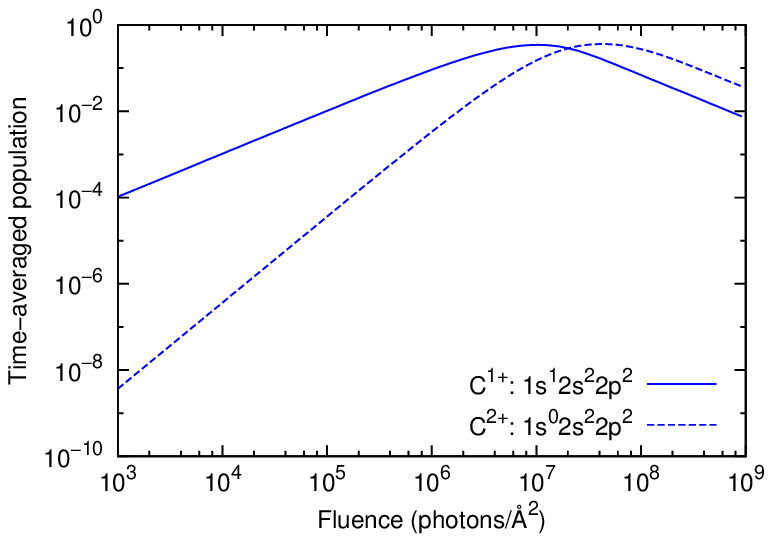}}
\caption{(Color online) Time-averaged populations of the single-core-hole and double-core-hole configurations for 8~keV and 12~keV photon energies.
The pulse length is fixed at 1~fs FWHM.}
\label{fig:prob}
\end{figure*}

As one can see in Fig.~\ref{fig:charge}, $\bar{Z}$ increases in increments of decreasing magnitude as the fluence becomes
higher, indicating a saturation effect.
To make this point clearer, we plot the time-averaged population $\bar{P}_I$ of the single ($I$=\config{1}{2}{2}) and double ($I$=\config{0}{2}{2}) core-hole configurations as a function of the fluence in Fig.~\ref{fig:prob}.
The pulse length is fixed at 1~fs FWHM, which is less than the lifetimes of the core-hole configurations.
Both populations follow a power-law dependence for fluences up to about 10$^6$ photons/\AA$^2$: single-core-hole
production is a one-photon process and therefore is a linear function of the fluence (below 10$^6$ photons/\AA$^2$),
whereas double-core-hole production is a two-photon process and therefore is a quadratic function of the fluence.
Saturation occurs around 10$^6$--10$^8$ photons/\AA$^2$.  For even higher fluences, the populations of both configurations
decrease due to further photoionization (core ionization in the single-core-hole configuration and valence ionization 
in the double-core-hole configuration, respectively).

\subsection{Influence of hollow-atom formation on x-ray scattering intensity}
\begin{figure*}
\centering
\subfigure[\ 8~keV]{\includegraphics[width=\figurewidth]{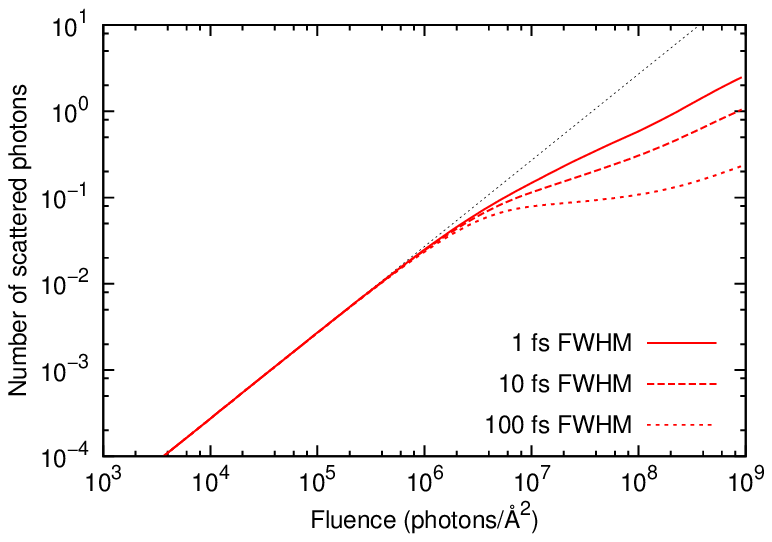}}
\subfigure[\ 12~keV]{\includegraphics[width=\figurewidth]{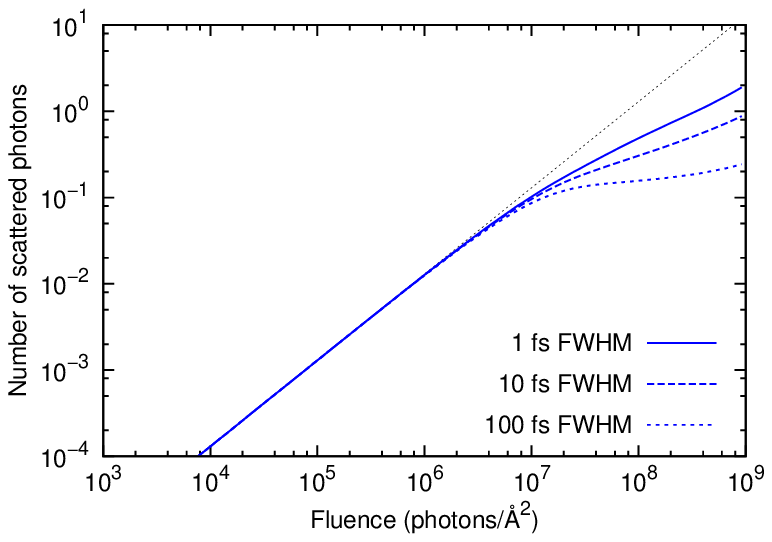}}
\caption{(Color online) Number of scattered photons for 8~keV and 12~keV photon energies.  
A spatial resolution of 1.7~\AA\ is assumed.
The thin dotted lines extrapolate the linear fluence dependence valid at low fluence.}
\label{fig:N_S}
\end{figure*}

The pulse-length and fluence dependence of the time-averaged charge affects the number of scattered photons, which must
be maximized in single-shot experiments such as to obtain an optimal signal-to-noise ratio.
In Fig.~\ref{fig:N_S}, the number of scattered photons is plotted as a function of the fluence for 8~keV and 12~keV photon energies.
The spatial resolution is fixed at $d$=1.7~\AA\ and three different pulse lengths are considered (1, 10, and 100~fs).
For fluences below $\sim 10^6$~photons/\AA$^2$, $N_\text{S}$ depends linearly on the fluence of incident photons, 
but is independent of the pulse length (see Fig.~\ref{fig:N_S}).  
In this low-fluence regime, the number of photons scattered per atom and per pulse is less than 0.1.  
In order to scatter at least 0.1 photons, the fluence must be in the regime above $10^6$~photons/\AA$^2$.  At high fluence, after saturation of inner-shell ionization, $N_\text{S}$ is no longer a linear function of the fluence and, particularly, depends sensitively on the pulse length.  
As can be seen in Fig.~\ref{fig:N_S}, the number of photons scattered may be maximized at a given fluence by using a pulse duration shorter than the double-core-hole lifetime ($\tau_\text{D} \approx 3$~fs).

If we require that a carbon atom scatters, say, 0.1 photons per pulse and per pixel, then, assuming a pulse length of 1~fs and a photon energy of 12~keV, a fluence of $1\times 10^{7}$ photons/\AA$^2$ per pulse is needed.  
However, if we assume a 10 fs pulse instead, then the fluence required would increase by a factor of four.
Figure~\ref{fig:N_S} illustrates quite distinctly the impact of hollow-atom formation on coherent x-ray scattering at high intensity.
In a molecule consisting of $N_\text{atom}$ atoms, the number of scattered photons is proportional to at least $N_\text{atom}^{1/3}$ per pulse and per pixel~\cite{Shneerson08}.
For example, with the above x-ray parameters, a molecule consisting of 100,000 carbon atoms would scatter at least 5 photons per pulse and per pixel.
Note that 5 photons per pixel would be sufficient for successful 3-dimensional structural reconstruction~\cite{Shen04}.

\subsection{Dependence of the $R$-factor on the desired resolution}\label{sec:resolution}

In addition to the strength of the scattering signal, another important factor is the quality of the x-ray scattering pattern.
The scattering pattern from the damaged sample should be as similar as possible to the scattering pattern that would be obtained if the sample were unaffected by radiation damage.
Using the $R$-factor in Eq.~(\ref{eq:R-factor}), we measure the quality of the x-ray scattering pattern.

\begin{figure}
\centering
\includegraphics[width=\figurewidth]{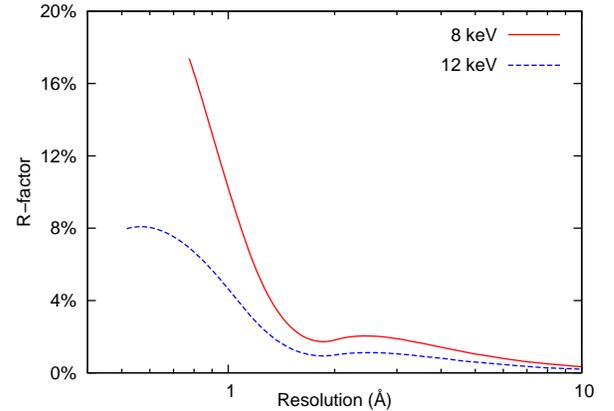}
\caption{(Color online) $R$-factor as a function of the spatial resolution for 8~keV and 12~keV photon energies.
The fluence is $10^{7}$~photons/\AA$^2$ and the pulse length is 1~fs FWHM.}
\label{fig:resolution}
\end{figure}               

In Fig.~\ref{fig:resolution}, we examine the correlation between the $R$-factor and the desired spatial resolution for 8~keV and 12~keV photon energies.  
The pulse length assumed is 1~fs FWHM and the fluence is fixed at $10^{7}$~photons/\AA$^2$.  
Under these conditions, inner-shell ionization is saturated (see Fig.~\ref{fig:prob}), and every ten pulses about one photon is scattered per carbon atom (see Fig.~\ref{fig:N_S}).
Each curve in Fig.~\ref{fig:resolution} ends at the finest spatial resolution possible at the respective photon energy.
Because of the reduction of electronic damage at higher photon energies, the $R$-factor at 12~keV is less than that at 8~keV.  
For a spatial resolution $d > 1$~\AA, the spatially localized reduction of electron density in the 1s shell in the single-core-hole and double-core-hole configurations is difficult to resolve, rendering the $R$-factor rather low.
The small local maximum around 2.5~\AA\ is due to the 2s vacancy formed by valence ionization in the double-core-hole configuration.
For very fine resolution ($d < 1$~\AA), the core vacancy can be resolved, so the $R$-factor rapidly increases.  
Note, however, that the $R$-factor values in Fig.~\ref{fig:resolution} are still less than 20\%, which is a typical value for x-ray crystallographic data~\cite{Hau-Riege05}.
For $d$=1.7--1.9~\AA, the $R$-factor takes on a local minimum value of less than 2\%.

\subsection{Role of impact ionization}\label{sec:impact_ionization}

So far, we have focused on electronic damage processes that are not strongly affected by the molecular environment.
An important damage mechanism characteristic of extended molecular systems is impact ionization by (quasi-)free electrons~\cite{Hau-Riege04,Jurek04,Ziaja05,Ziaja06,Kai10}.  
For an x-ray pulse much shorter than inner-shell decay lifetimes, impact ionization by Auger electrons is irrelevant for the formation of electronic damage during the pulse.
On the other hand, impact ionization by photoelectrons is not, in general, negligible.
Here we discuss how to reduce photoelectron impact ionization by using short pulses.

\begin{figure}
\centering
\includegraphics[width=\figurewidth]{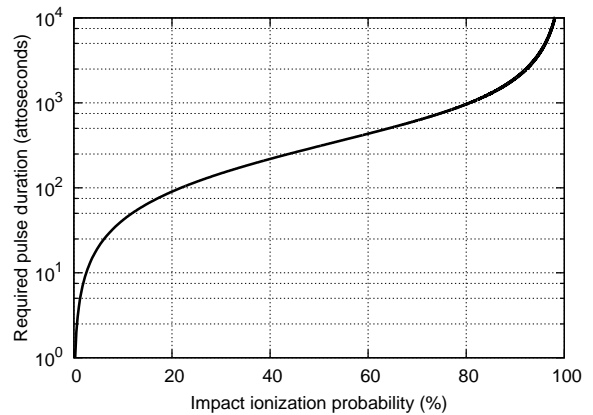}
\caption{Plot of the pulse duration required for a given impact ionization probability, for a photoelectron with a kinetic energy of 12~keV in a carbon-based medium with a mean free path of 13~nm.}
\label{fig:impact_ionization}
\end{figure}

The mean free path for a 12~keV photoelectron in a carbon-based medium (diamond) is about 13~nm~\cite{Ziaja06}.
Let us assume that the photoelectron travels in the $x$ direction in a homogeneous sample.
From the definition of the mean free path, it follows that the impact ionization probability is given by
\begin{equation}
P_\text{impact}(x) = 1 - e^{-x / \lambda},
\end{equation}
where $x$ is the distance traveled and $\lambda$ is the mean free path.
If we allow an impact ionization probability of 20\%, the maximum $x$ permitted is $-\lambda \log(1-0.2) = 22\% \times \lambda$.
Therefore, for molecules with a diameter of 3~nm or less, the impact ionization probability per photoelectron is less than 20\%.
For much larger molecules, the role of impact ionization can be reduced by using an x-ray pulse that is so short that it is over before impact ionization has taken place with a probability less than a certain percentage.
In analogy to Eq.~(\ref{eq:time-averaged_population}), one can define the time-averaged impact ionization probability during the x-ray pulse as
\begin{equation}
\bar{P}_\text{impact} = \int_0^\tau P_\text{impact}\left( x(t) \right) f(t) \; dt,
\end{equation}
where $\tau$ is the pulse duration and $f(t)$ is the normalized x-ray pulse envelope.
It is assumed that the electron starts to travel at the beginning of the pulse, i.e., $x(t) = v t$, where $v$ is the speed of the photoelectron.
If we use a flat-top pulse envelope, $f(t) = 1 / \tau$ for $0 \leq t \leq \tau$, then $\bar{P}_\text{impact} = 1 + \lambda / ( v \tau ) [ \exp( -v \tau / \lambda ) - 1 ]$.
Figure~\ref{fig:impact_ionization} plots $\tau$ versus $\bar{P}_\text{impact}$ for a mean free path of 13~nm and a photoelectron energy of 12~keV.
For a pulse-weighted impact ionization probability of 20\%, the pulse duration required is about 100 attoseconds, corresponding to a Fourier-limited bandwidth of the order of 10~eV.  
In calculations using a Gaussian pulse envelope, we obtained very similar results.

\section{Conclusion}\label{sec:conclusion}
In this paper, we have investigated electronic damage and coherent x-ray scattering using ultrashort and ultraintense x-ray pulses attainable with current and future x-ray FELs.  
For all possible electronic configurations of the atomic system, we have calculated rate parameters for x-ray-induced damage processes including photoionization, Auger decay, and fluorescence, in a consistent \textit{ab initio} framework.  
The impact of electronic damage has been studied by employing a set of coupled rate equations, which we have incorporated into simulations of coherent x-ray scattering signals.
We have implemented an integrated toolkit, \textsc{xatom}, to treat all above-mentioned processes.

Our numerical simulations of coherent x-ray scattering signals including electronic damage dynamics show that hollow-atom formation and the associated phenomenon of x-ray transparency or frustrated absorption play a crucial role in optimizing the strength and quality of single-shot x-ray scattering signals.
Hollow-atom formation is particularly important when the x-ray pulse length is a few femtoseconds or shorter, and saturates, in the case of carbon, around a fluence of 10$^6$--10$^8$~photons/\AA$^2$, corresponding to 10$^{12}$--10$^{14}$ photons per pulse at a beam size of 100~nm$\times$100~nm.  
At a fluence of 10$^7$~photons/\AA$^2$, for instance, the number of photons scattered per pulse, within a spatial resolution of 1.7~\AA\, is about 0.1 per carbon atom, at a pulse length of 1~fs and a photon energy of 12~keV.  
A hollow atom is resistant to further electronic damage via photoionization and, for a spatial resolution $d > 1$~\AA, gives rise to an x-ray scattering pattern that differs little from that obtained for the neutral atom in its ground electronic configuration.
A comparison between our data for 8~keV and 12~keV photon energies shows that there are no qualitative differences.
By using a higher photon energy, the quality of the scattering pattern, as defined by the $R$-factor, can be increased (by reducing electronic damage), but the number of photons scattered per pulse decreases somewhat.

Finally, we have analyzed the role of impact ionization in molecules and provided a simple estimate of the pulse duration required to suppress impact ionization during the x-ray pulse.
This estimate, in combination with the calculations presented in this paper, suggests that attosecond x-ray FELs~\cite{Zholents04,Saldin06,Zholents08,Ding09} with a pulse length of $\sim 100$~as, $\sim 10^{13}$ photons per pulse, and a photon energy of $\sim 12$~keV are ideal for single-shot imaging of individual macromolecules at atomic resolution.

\begin{acknowledgments}
We thank Stefan Pabst for helpful discussions.
This work was partially supported by the Chemical Sciences, Geosciences, and Biosciences Division of the Office of Basic Energy Sciences, Office of Science, US Department of Energy (DE-AC02-06CH11357).
\end{acknowledgments}

%

\end{document}